\documentclass[journal]{IEEEtran}
\usepackage{graphicx}

\begin{document}
\title{Commissioning of the Testbeam Prototype\
of the CALICE Tile Hadron Calorimeter}

\author{~Benjamin Hermberg on behalf of the CALICE collaboration
     
\thanks{Manuscript received November 20, 2012.}
\thanks{B. Hermberg is with DESY, Hamburg, Germany (email: benjamin.hermberg@desy.de)}%
}

\maketitle
\pagestyle{empty}
\thispagestyle{empty}

\begin{abstract}
The CALICE collaboration is currently developing an engineering prototype of an analog hadron calorimeter for a future linear collider like the ILC. One main task of this prototype is to demonstrate the feasibility of building a realistic detector with fully integrated front-end electronics. To reach the required high jet energy resolution the detector is based on highly segmented read out layers. The signals are sampled by scintillating tiles that are individually read out by silicon photomultipliers. The readout chips are integrated into the calorimeter layers and are optimized for minimal power consumption. An LED system is integrated into each channel of the detector to calibrate the photodetectors. This report presents the realization of a setup of one layer and the performance in a testbeam  environment.
\end{abstract}

\section{Introduction}

\begin{figure}[!t]
\centering
\includegraphics[width=3.5in]{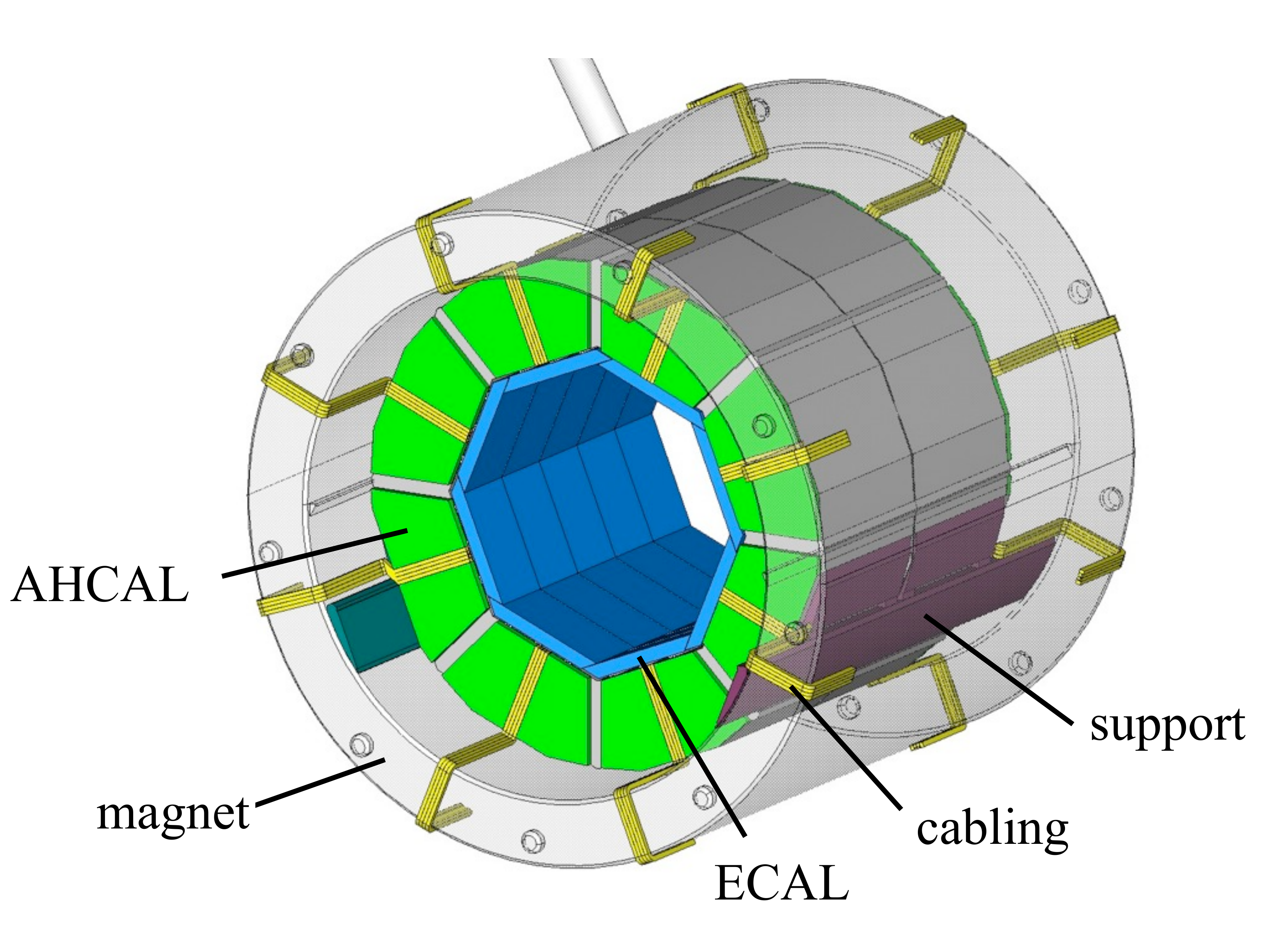}
\caption{Layout of the barrel calorimeter system of a LC detector  ~\cite{Reinecke}. The AHCAL is shown in green and the ECAL in blue, while the structure is surrounded by the magnet.}
\label{fig_AHCAL}
\end{figure}

\begin{figure}[!t]
\centering
\includegraphics[width=3.5in]{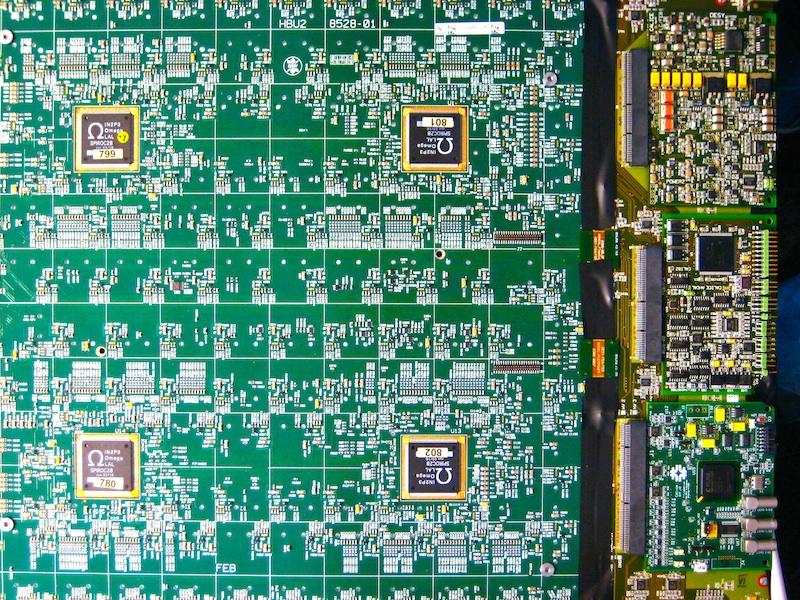}
\caption{Photo of one single HBU connected to the CIB (central interface board). The CIB hosts the power board, the calibration board (Calib) and the detector interface board (DIF) (top to bottom).}
\label{fig_HBU}
\end{figure}

\IEEEPARstart{W}{ithin}  the CALICE collaboration~\cite{CALICE} new technologies for calorimeters for a future electron - positron collider experiment are currently under development. Figure 1 shows a possible design of a barrel calorimeter for such a detector.
The analog hadron calorimeter will be build as a sampling calorimeter with 48 layers in a cylindrical structure. The inner and outer radius of the calorimeter are $2.0\,$m and $3.1\,$m. Inside of the analog hadron calorimeter (AHCAL) the electromagnetic calorimeter (ECAL) will be placed. The calorimeter system is surrounded by the magnet (see figure~\ref{fig_AHCAL}). To improve the jet resolution compared to previous experiments a good shower separation combined with the information from the tracker system is used. This approach is known as particle flow concept~\cite{AdloffI} and has been validated with the physics prototype of the CALICE AHCAL~\cite{AdloffII}. A very high segmentation of the calorimeters in all three spatial dimensions is mandatory to guarantee a good performance of the particle flow algorithm. A new engineering prototype~\cite{Reinecke} is currently being developed to demonstrate that a scalable device can be built that meets the requirements of linear collider (LC) experiments. Key issues for the readout electronics are the low power consumption to avoid an active cooling and the full integration into the active calorimeter layers to minimize dead zones. The prototype is based on scintillating tiles that are read out by silicon photomultipliers. Each detector subunit (HCAL base unit, HBU) comprises 144 detector channels and has a size of 36$\times$36$\,$cm$\textsuperscript{2}$ (see figure~\ref{fig_HBU}). Several HBUs have been extensively tested in the laboratory as well as in an electron beam at the DESY testbeam facility. The concept, the status of the prototype and results from testbeam measurements are presented here.     

\section{Concept and status of the engineering prototype}

The barrel of the hadron calorimeter is divided in two parts along the beam axis, each part is segmented in 16 sectors in $\phi$ direction. Each sector has 48 layers. One layer consists of the tiles, the embedded front-end electronics and a 16$\,$mm thick stainless steel or 10$\,$mm thick tungsten absorber plate. The total number of channels in the hadron calorimeter barrel adds up to 3.9 million. For easy maintenance and service all of the electronics connections and interface modules are placed at the two end-faces of the barrel.

\subsection{Detector, Tiles and ASICs}

One active layer consists of three parallel slabs. Each slab is subdivided in six HBUs,  interconnected via ultra thin flex-leads. The middle slab is connected to the DAQ via the Central Interface Board (CIB) and the side slabs are in turn connected to the CIB via Side Interface Boards (SIBs). The scintillating tiles which are connected to the channels with a size of 3$\times$3$\times$0.3$\,$cm$\textsuperscript{3}$ are sampling the deposited energy in the calorimeter. The scintillating light is guided with an integrated wavelength shifting fiber to a Silicon Photomultiplier (SiPM) with a size of 1.27$\,$mm$\textsuperscript{2}$. The SiPM comprises 796 pixels operated in Geiger mode with a gain of $\sim$$\,$0.5$\,$$\cdot$$\,$10$\,$$\textsuperscript{6}$ -- 2.0$\,$$\cdot$$\,$10$\,$$\textsuperscript{6}$. 

The tiles have nominal distances of $\sim\,$100$\,\mu$m and are connected to the HBUs printed circuit board (PCB) with two alignment pins by plugging them into holes in the PCB. The analog signals of the SiPMs are read out by four 36$\,$-$\,$channel ASICs (SPIROC2b)~\cite{Raux} equipped with 5$\,$V DACs for a channel-wise bias voltage adjustment on each HBU (offering a coverage of $\sim\,$10$\,\%$ of the operation voltage, see figure~\ref{fig_bias}). In order to set the bias voltage for each SiPM very precisely to ensure a correct quantum efficiency a high resolution DAC response curve is measured for each SiPM (see figure~\ref{fig_input}), because the DAC output characteristic varies from channel to channel.\\
The ASICs are lowered into cutouts by $\sim\,$500$\,\mu$m to reduce the height of the active layers. Each channel offers two gain modes to cover a dynamic range of 1 to 2000 photoelectrons, where the high gain mode is primarily foreseen for taking calibration data and the low gain mode to measure signals with higher amplitudes up to SiPM saturation. To avoid the need for an active cooling system between the calorimeter layers it is foreseen to reduce the power consumption to 25$\,\mu$W (40$\,\mu$W including SiPMs) per channel for the final LC operation. This can be reached only by switching off the unused parts of the ASICs between the bunch trains (power cycling). The on-detector zero suppression with an adjustable threshold is integrated into the ASICs as well as the digitization step with a 12$\,$-$\,$bit ADC (analog to digital converter) for charge and a 12$\,$-$\,$bit TDC (time to digital converter) for time measurements. The TDC comprises two ramps with variable lengths of 200$\,$ns and 5$\,\mu$s, depending on the operation mode (LC or testbeam mode). 

\begin{figure}[!t]
\centering
\includegraphics[width=3.5in]{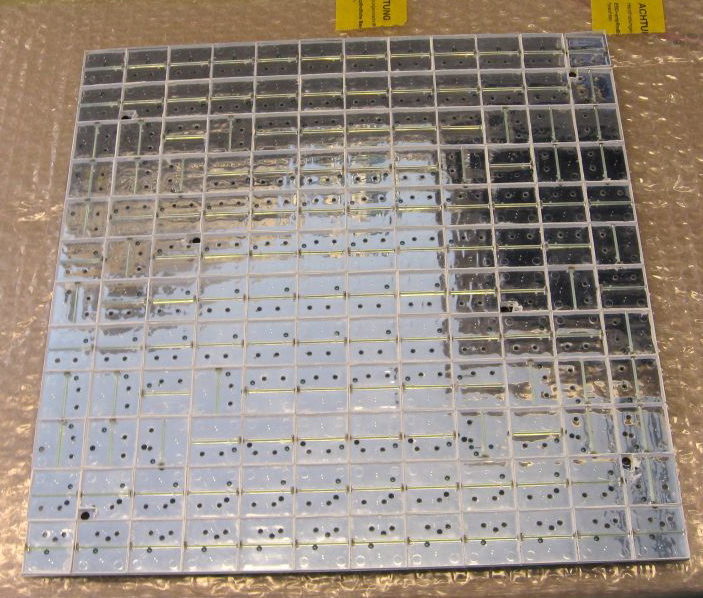}
\caption{One HBU fully equipped with tiles. The orientation of the wavelength shifting fibers depends on the holes for the pins. This is determined by the details of the PCB design.}
\label{fig_tiles}
\end{figure}

\begin{figure}[!t]
\centering
\includegraphics[width=3.5in]{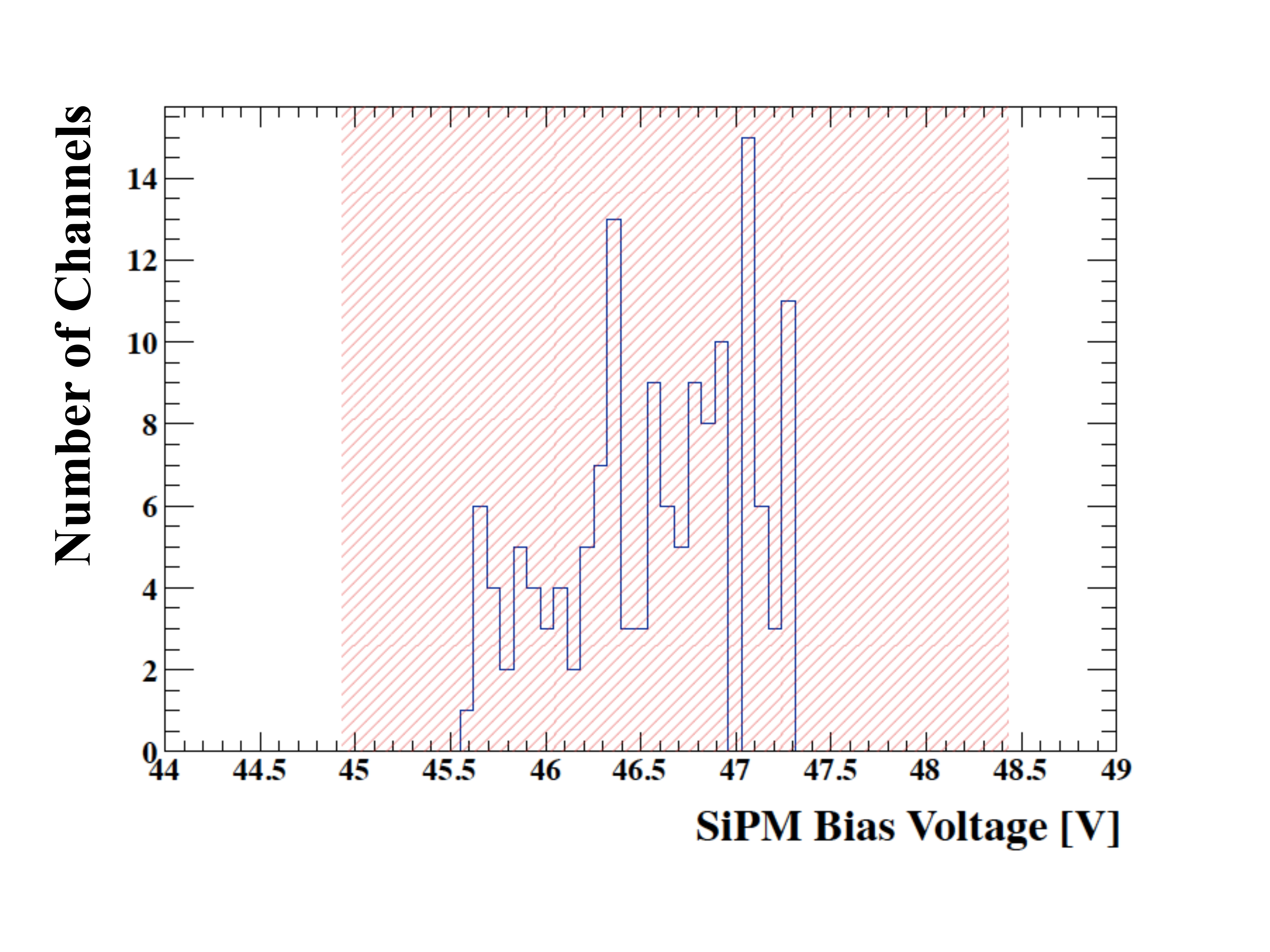}
\caption{Distribution of the SiPM bias voltages on one HBU. The hatched area displays the coverage of the input DACs for individual bias voltage adjustment~\cite{hartbri}.}
\label{fig_bias}
\end{figure}

\begin{figure}[!t]
\centering
\includegraphics[width=3.5in]{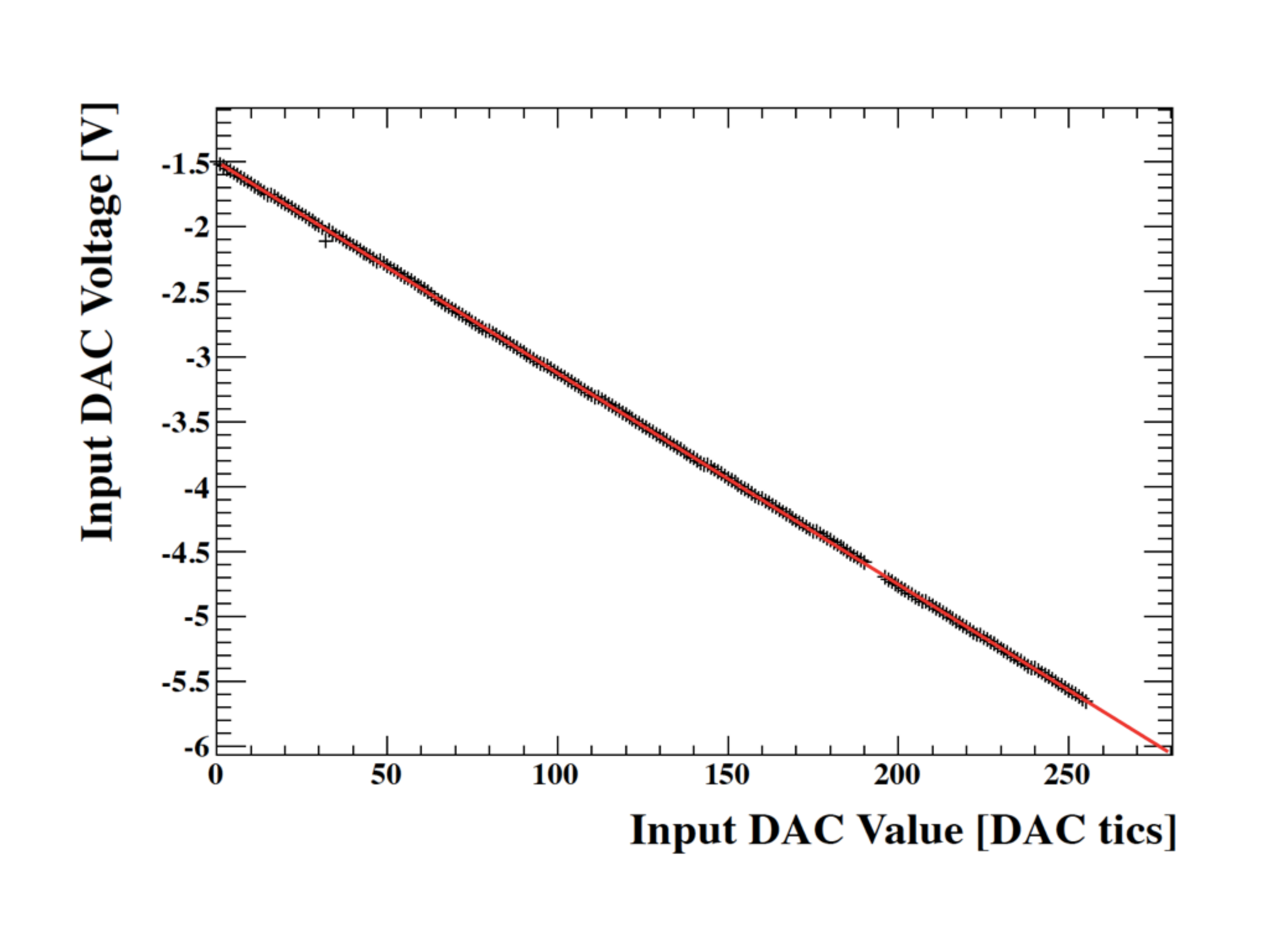}
\caption{High resolution input DAC response curve of a single channel~\cite{hartbri}.}
\label{fig_input}
\end{figure}

\subsection{Light Calibration System}

Since the SiPM response depends on the temperature (\mbox{$\sim\,$-1.7$\,\%$/K}), the bias voltage ($\sim\,$2.5$\,\%$/100$\,$mV) and the saturation due to the limited numbers of pixels, a gain- and saturation- monitoring system with a high dynamic range is needed to correct these effects. Two calibration and monitoring concepts are currently under development:
\begin{itemize}
\item Each channel contains a circuit for one integrated UV-LED that shines light on the corresponding tile. This system is used in the HBUs in the DESY test setups.
\item One strong LED is placed outside of the detector on a special interface board, where the light is distributed via notched fibers into the tiles. 
\end{itemize}
Both options have been successfully tested on the DESY setups in the laboratory and under testbeam conditions. In the calibration mode of the ASICs a very low light intensity is used to determine the gain from the distance between the peaks in a single pixel spectrum~\cite{Reinecke}. At higher light intensities (corresponding to $\sim\,$100 minimum ionizing particles (MIPs)) the SiPMs show saturation behavior.

\subsection{Current Setup and DAQ} 

\begin{figure}[!t]
\centering
\includegraphics[width=3.5in]{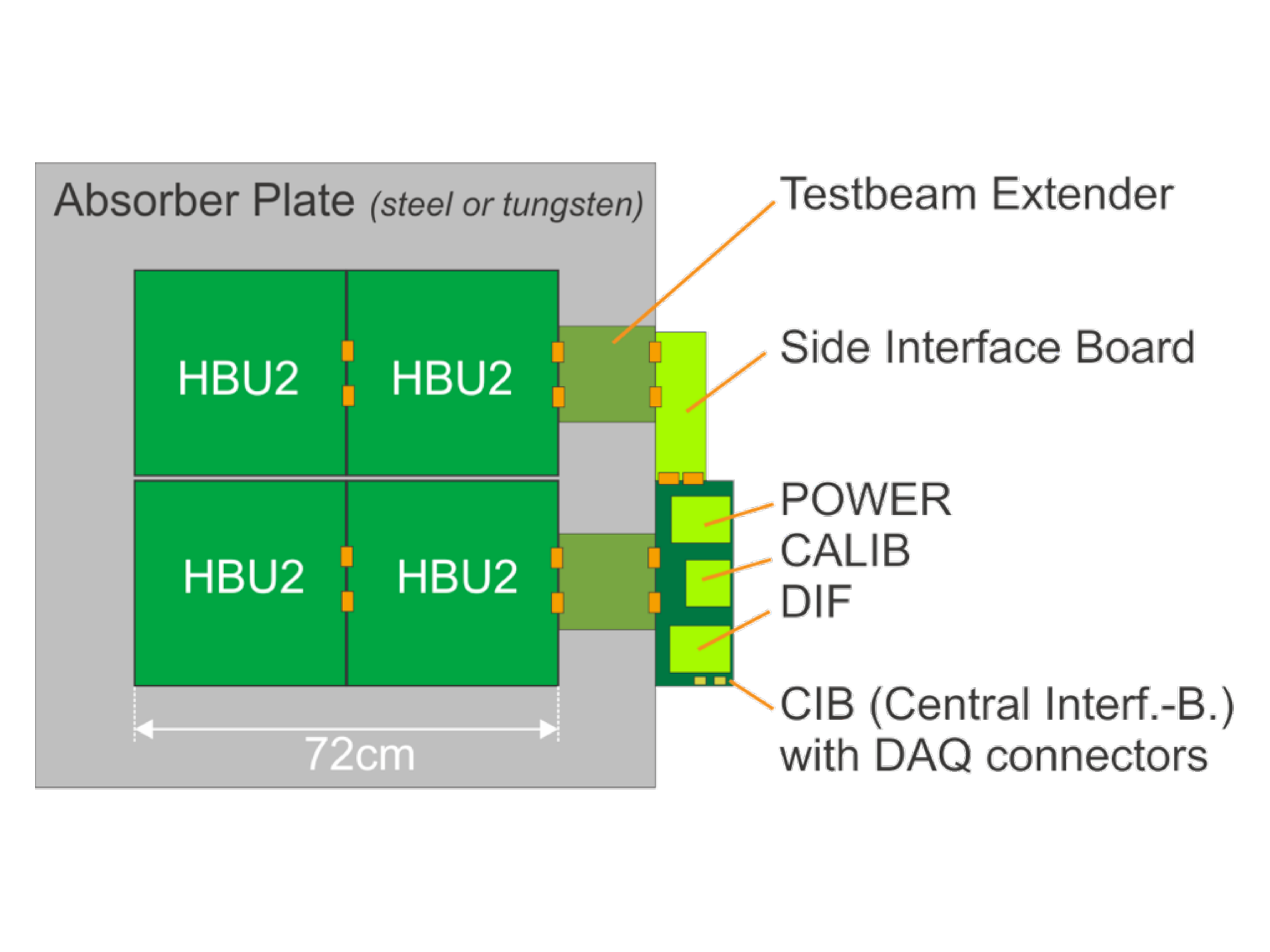}
\caption{Schematic of the large layer setup with four HBUs as it is foreseen for a hadron testbeam at a CERN facility.}
\label{fig_testbeam_setup}
\end{figure}

\begin{figure}[!t]
\centering
\includegraphics[width=3.5in]{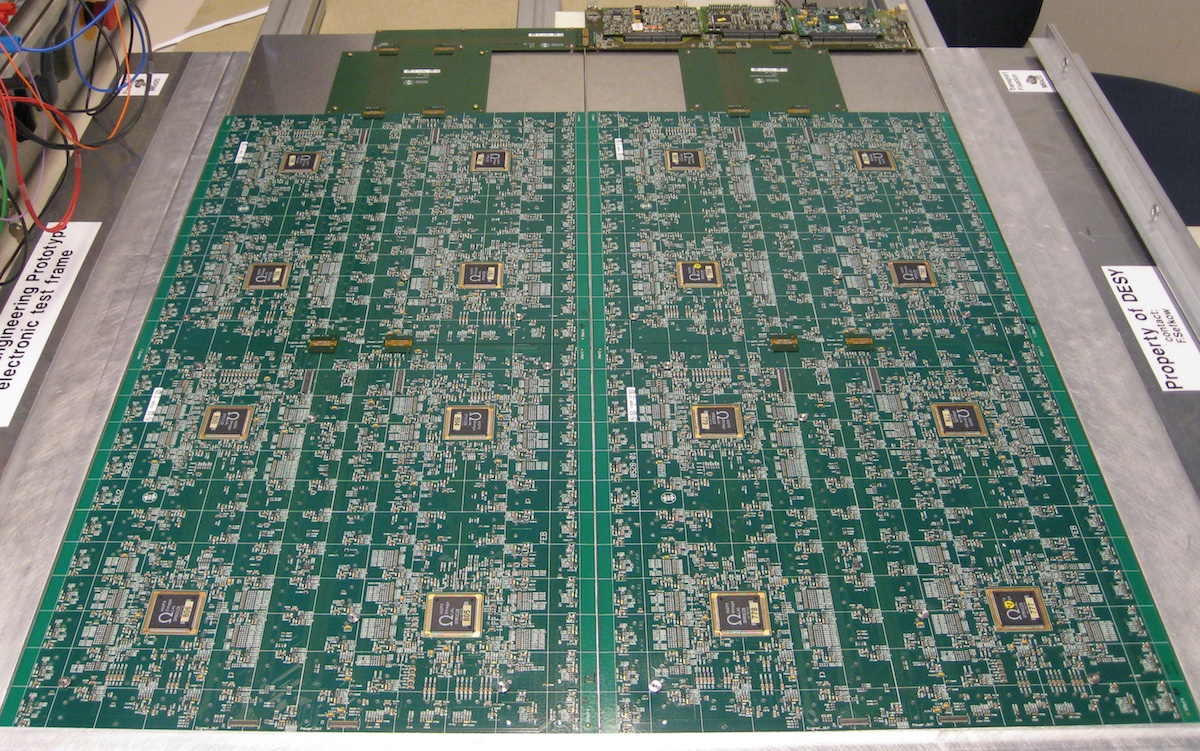}
\caption{Photo of the large layer setup.}
\label{fig_full_layer}
\end{figure}

The current setup consists of two slabs with two HBUs (figures~\ref{fig_testbeam_setup} and~\ref{fig_full_layer}). These four HBUs build a layer setup which can be integrated into a detector. Altogether the layer setup consists 576 read out channels with a size of 72$\times$72$\,$cm$\textsuperscript{2}$. In order to fit into current HCAL testbeam absorber structures of 1$\times$1$\,$m$\textsuperscript{2}$ the testbeam extender boards have to be installed to have both the HBUs centered and the DIF located outside as shown in figure~\ref{fig_testbeam_setup}. The first slab comprises the Central Interface Board that hosts the Detector Interface, the steering board for the calibration system and the power module, which distributes all voltages needed in the slab. The second slab is connected to the DAQ via the Side Interface Board. 

\section{Measurements and Results}

The main goal of the current tests of the prototype modules is the commissioning of a multi HBU prototype. On one hand this requires tests of the functionality and performance of all subcomponents in the laboratory as well as in a testbeam environment. This has been done in the past and is still ongoing. Results of recent tests are reported in the following. On the other hand, system aspects have to be considered and the performance of the large setup has to be tested. Here the tests are also still ongoing, like a large system test in a testbeam environment at the CERN SPS testbeam facility in November 2012.  

\subsection{Channel Gain Equalization}

\begin{figure}[!t]
\centering
\includegraphics[width=3.5in]{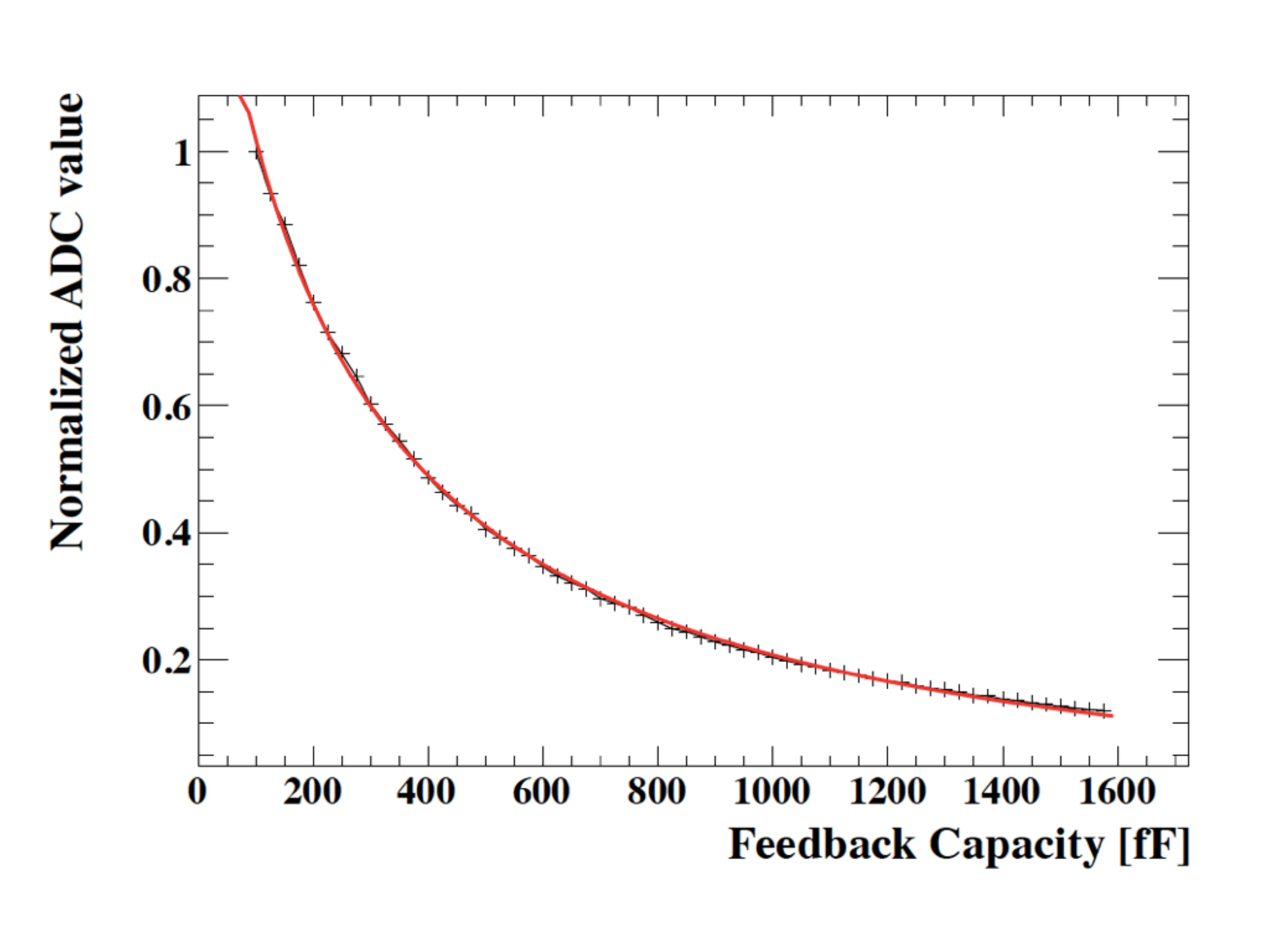}
\caption{Signal amplitude normalized to the signal measured with reference preamplifier setting (100$\,$fF) for different preamplifier settings in one channel~\cite{hartbri}.}
\label{fig_ref}
\end{figure}

\begin{figure}[!t]
\centering
\includegraphics[width=3.5in]{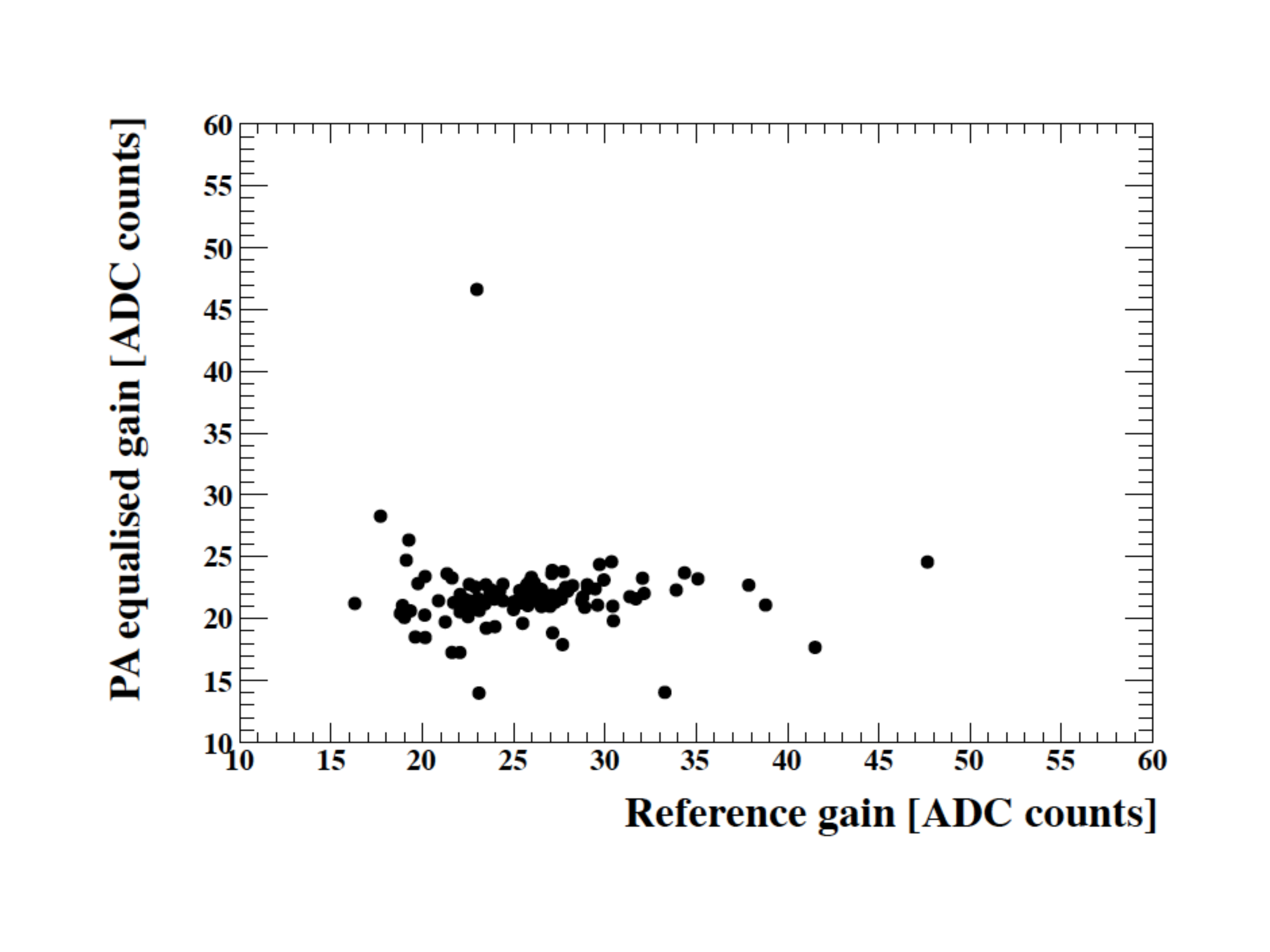}
\caption{Distribution of cell gain before and after the cell gain equalization. Some outliners are caused by a not usable fit of the gain dependence on the preamplifier settings in this channel ~\cite{hartbri}.}
\label{fig_equalize}
\end{figure}

To build a large prototype with four HBUs several calibration studies have to be done. Due to the production chain of the SiPMs, the properties of the SiPMs underly fluctuations with the result that the SiPMs have a wide gain spread. This would mean that the detector response varies from channel to channel for the same deposited energy in the scintillating tiles. To correct this effect the SiPM gains have to be equalized~\cite{hartbri}. The gain in each channel is a combination of the SiPM gain and a preamplifier setting. This combined gain is called cell gain C and is defined as C$\,$=$\,$N$\cdot$G, where G is the intrinsic SiPM gain and N the preamplifier conversion factor. While G varies with bias voltage U$_{bias}$ applied to the SiPM, changing of the bias voltage will also influence the photon conversion efficiency and therefore influences the light yield of the desired 15 pixels/MIP. To avoid changing the light yield, N can be set up for each channel. G varies in the range of 5$\cdot$10$\textsuperscript{5}$ to 2$\cdot$10$\textsuperscript{6}$, so N has to cover a range of around a factor 5 to compensate the variation in G.  For the gain calibration small amplitude light pulses are flashed into each scintillator tile. With the light pulses single pixel spectra (SPS) for each SiPM can be measured. Out of these SPS the cell gain for a fixed reference preamplifier setting can be extracted. After measuring the dependence of N on the preamplifier setting (see figure~\ref{fig_ref}) the cell gain can be adjusted to the target value for all cells. The validity of the procedure is verified by measuring the cell gain again with the new preamplifier settings. The spread of the gain before and after the equalization is shown in figure~\ref{fig_equalize}.      

In order to have a well calibrated large layer this procedure is done for all four HBUs and the original gain spread of up to a factor 6 can be compensated and a typical uniformity of 5-10$\,\%$ is reached. 

\subsection{Time Measurement}
\begin{figure}[!t]
\centering
\includegraphics[width=3.5in]{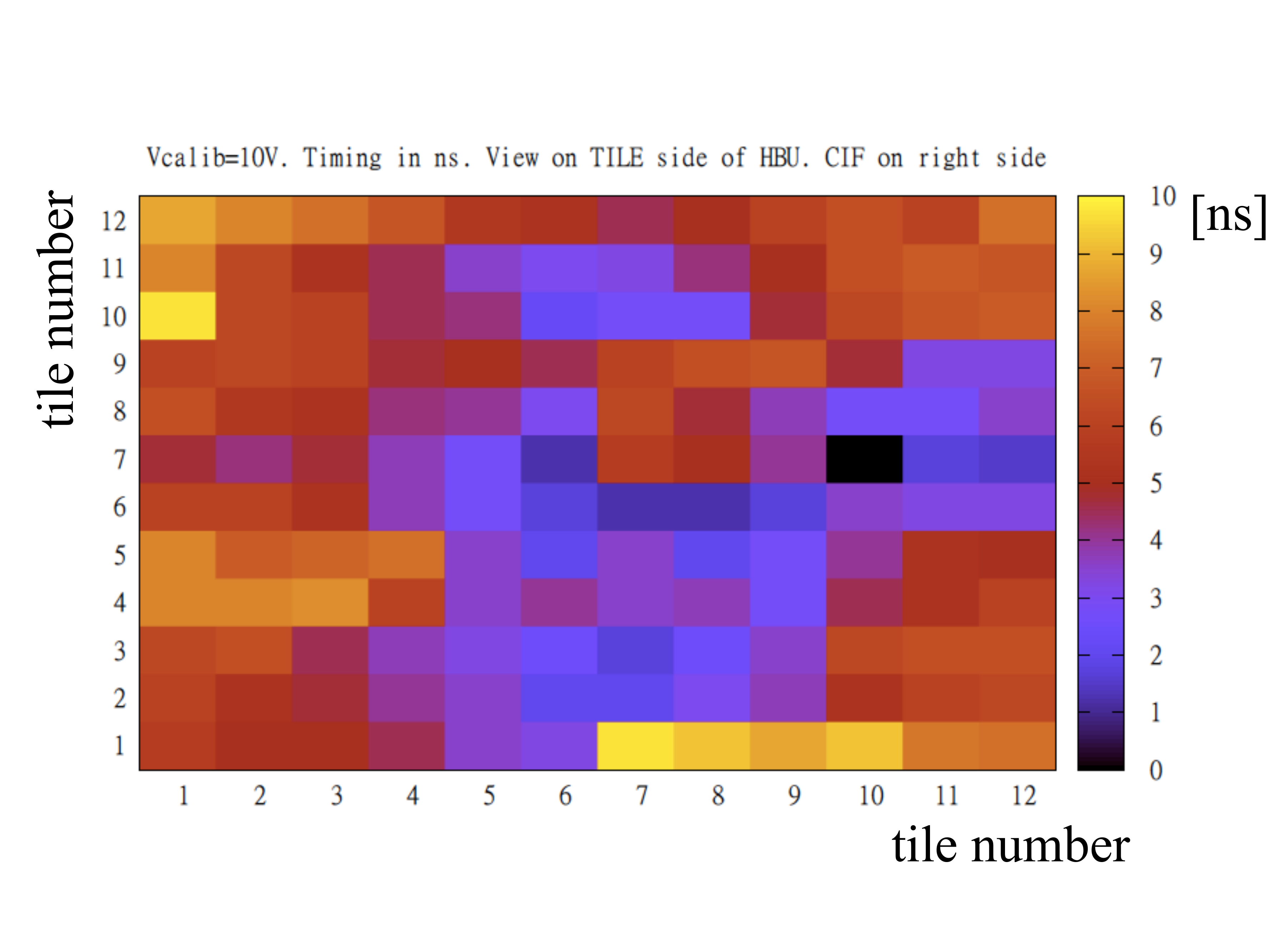}
\caption{Time delay between external trigger and LED pulse on one HBU~\cite{hartbri}.}
\label{fig_pattern}
\end{figure}

Besides the system tests of the prototype another main goal for the large layer test at the CERN testbeam facility is to study the time development of hadronic showers. 
In addition to the energy measurement of the shower the measurement of the arrival time of the signal in each cell is necessary. With this information it is possible to distinguish between prompt and delayed shower components, which are usually caused by late neutrons. One feature of the SPIROC2b is that it has the capability to determine the arrival time relative to the bunch clock of each event in each cell with a dual TDC ramp. The TDC uses a voltage ramp which is started with the rising edge of the clock and is reset with the next rising clock edge. Since the reset of the ramp introduces dead time, two ramps are implemented. The ramps are switches by a multiplexer (which also introduces some dead time). 
Two possible operation modes are foreseen in the current HBU design. In the ILC mode the ramp has a length of 200$\,$ns (which fits to the foreseen bunch structure of the ILC), while in testbeam mode the ramp has a length of about 5$\,\mu$s (in order to reduce the dead time due to multiplexing). To exploit the full dynamic range of the TDC in case of a change of the ramp length, the slope of the ramp can also be changed by bias points on the ASIC. The physics goal of the performance of the time measurement is to achieve a resolution of about 1 -- 3$\,$ns in testbeam mode to be able to distinguish between prompt and late shower components. With the current setup, TDC resolutions of $\sim\,$250$\,$ps in ILC Mode and 2 -- 3$\,$ns in testbeam mode are reached for single channels. These can be improved by increasing the slope to use the fall dynamic range , and in testbeam mode also by reducing the ramp length.\\ 
Also for the TDC measurement channel by channel variations are observed, which can be calibrated using the LED system. If the LED system is used to determine the offset of the TDC the time delay between an external trigger and an onset of LED emission has to be taken into account, caused by the spatial distribution of the LEDs on the HBU. The time delay for an HBU is shown in figure~\ref{fig_pattern}~\cite{hartbri} and the delay covers a range up to 9$\,$ns.\\    

\begin{figure}[!t]
\centering
\includegraphics[width=3.5in]{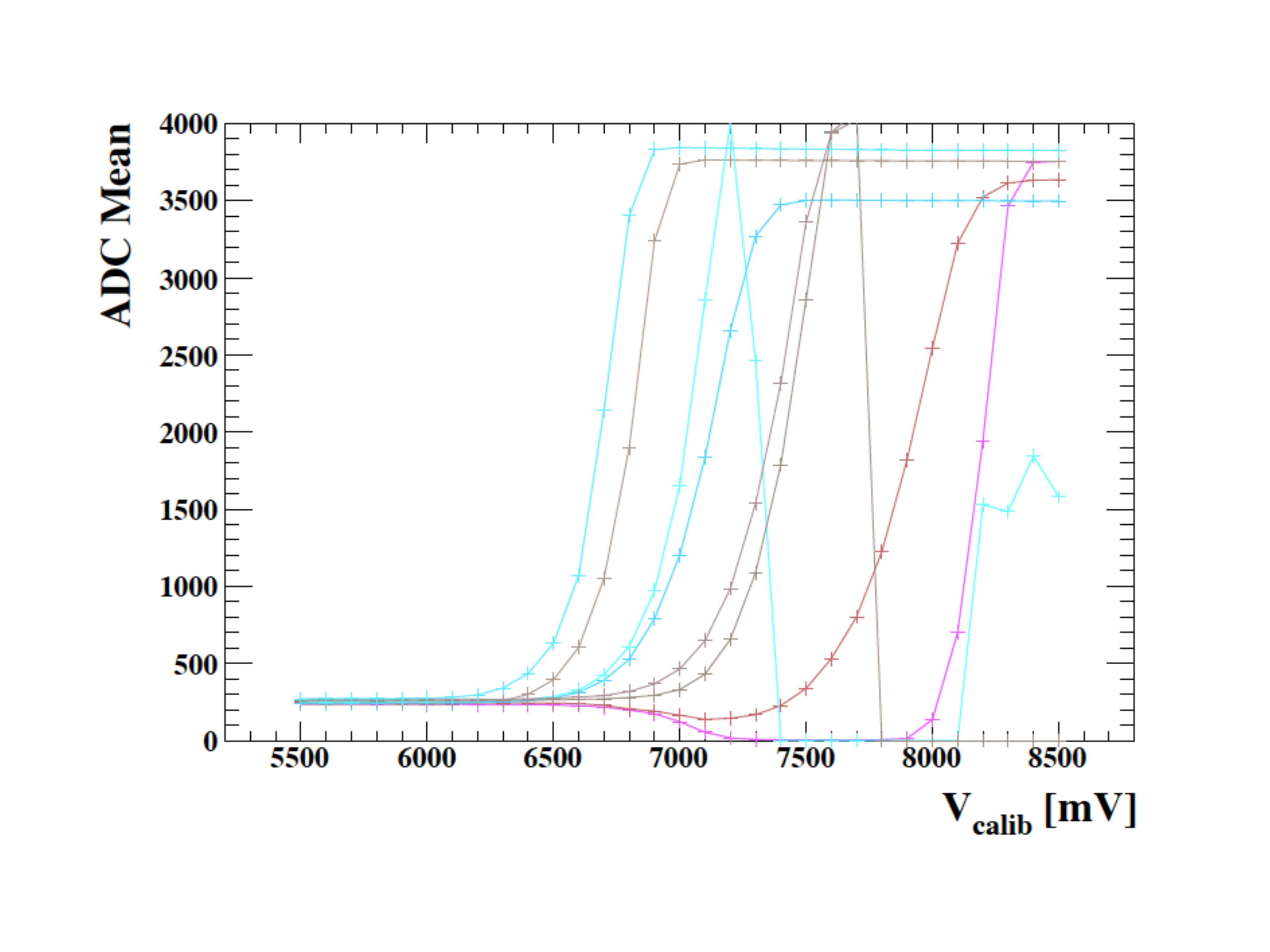}
\caption{LED response for several channels in ADC counts as a function of V$_{calib}$ voltage~\cite{hartbri}.}
\label{fig_Vcalib}
\end{figure}

\begin{figure}[!t]
\centering
\includegraphics[width=3.5in]{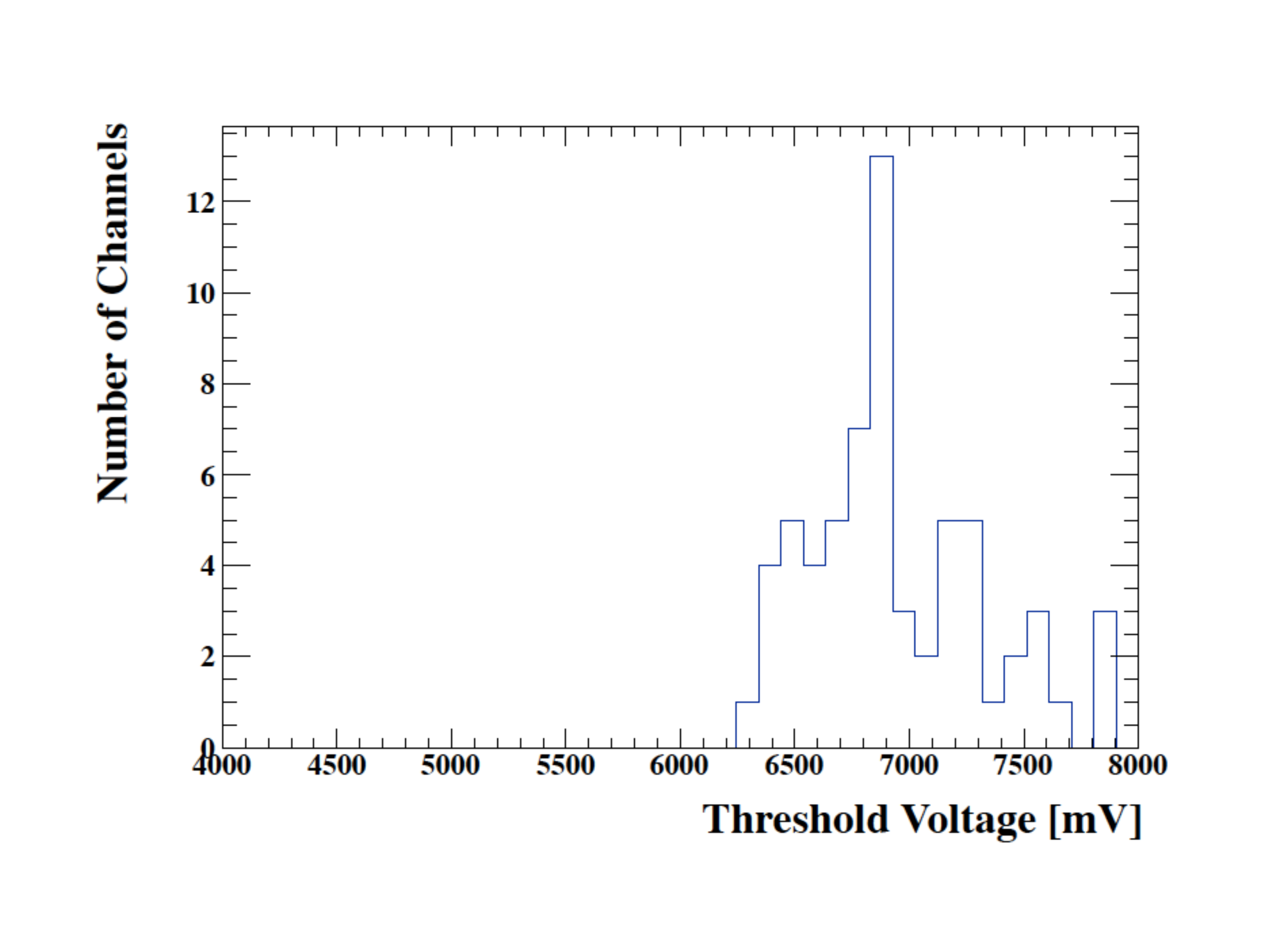}
\caption{Extracted voltage threshold distribution for the LEDs~\cite{hartbri}. }
\label{fig_threshold}
\end{figure}

\subsection{Light Amplitude Equalization of LED System and SiPM Saturation}

The main task of the light calibration system is the determine the gain of each SiPM by pulsing LED light of low intensity into the tiles. For practical reasons, this should be achieved in as few calibration runs as possible. Therefore the light output of the LEDs have to be equalized. The goal is to find a suitable combination and optimal values of three capacitors for each tile, such that the range of the LED bias voltages (V$_{calib}$) needed for the LED to produce photons is as small as possible. Each possible combination of capacitors has been used to measure the LED light output as a function of the bias voltage for a small set of tiles to equalize the light output of the LEDs. Nevertheless each LED needs a different bias voltage to start flashing.  Figure~\ref{fig_Vcalib} shows the LED response versus the bias voltage for different LEDs. Figure ~\ref{fig_threshold} shows the extracted voltage threshold distribution from figure~\ref{fig_Vcalib}. Nevertheless, further optimization of the capacity values is needed and further studies will be done in order to optimize the calibration procedure. The LED system is also used to measure the saturation of the SiPMs. Further information is given in~\cite{hartbri}.

\subsection{Calibration with MIPs}
Each HBU is used separately in an electron beam of 2$\,$GeV to investigate the response of the system to MIPs and to use that information to calibrate a large layer setup for a hadron testbeam run in November 2012 at the CERN SPS testbeam facility. For this purpose the HBUs are enclosed into light tight aluminum cassettes and mounted on a movable stage in order to scan all channels individually and separately. The MIP signals are measured in auto-trigger mode, where the threshold is optimized for each channel to measure a full MIP spectrum, while suppressing most of the SiPM noise. 
\begin{figure}[!t]
\centering
\includegraphics[width=3.5in]{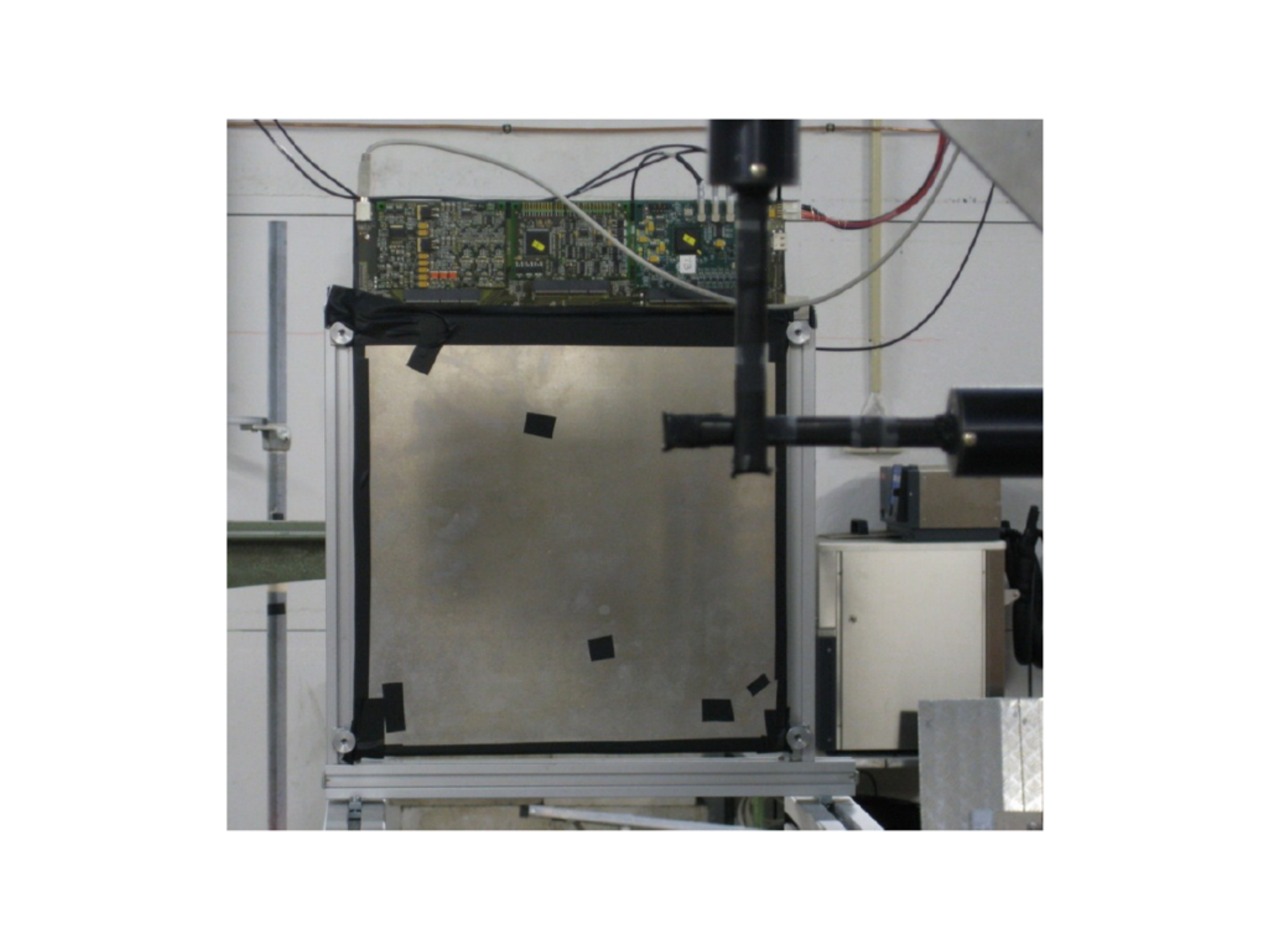}
\caption{Setup of the light tight HBU cassette as it is mounted on a movable stage at the DESY testbeam facility. On top of the cassette the detector interface modules are visible.}
\label{fig_testbeam}
\end{figure}

\begin{figure}[!t]
\centering
\includegraphics[width=3.5in]{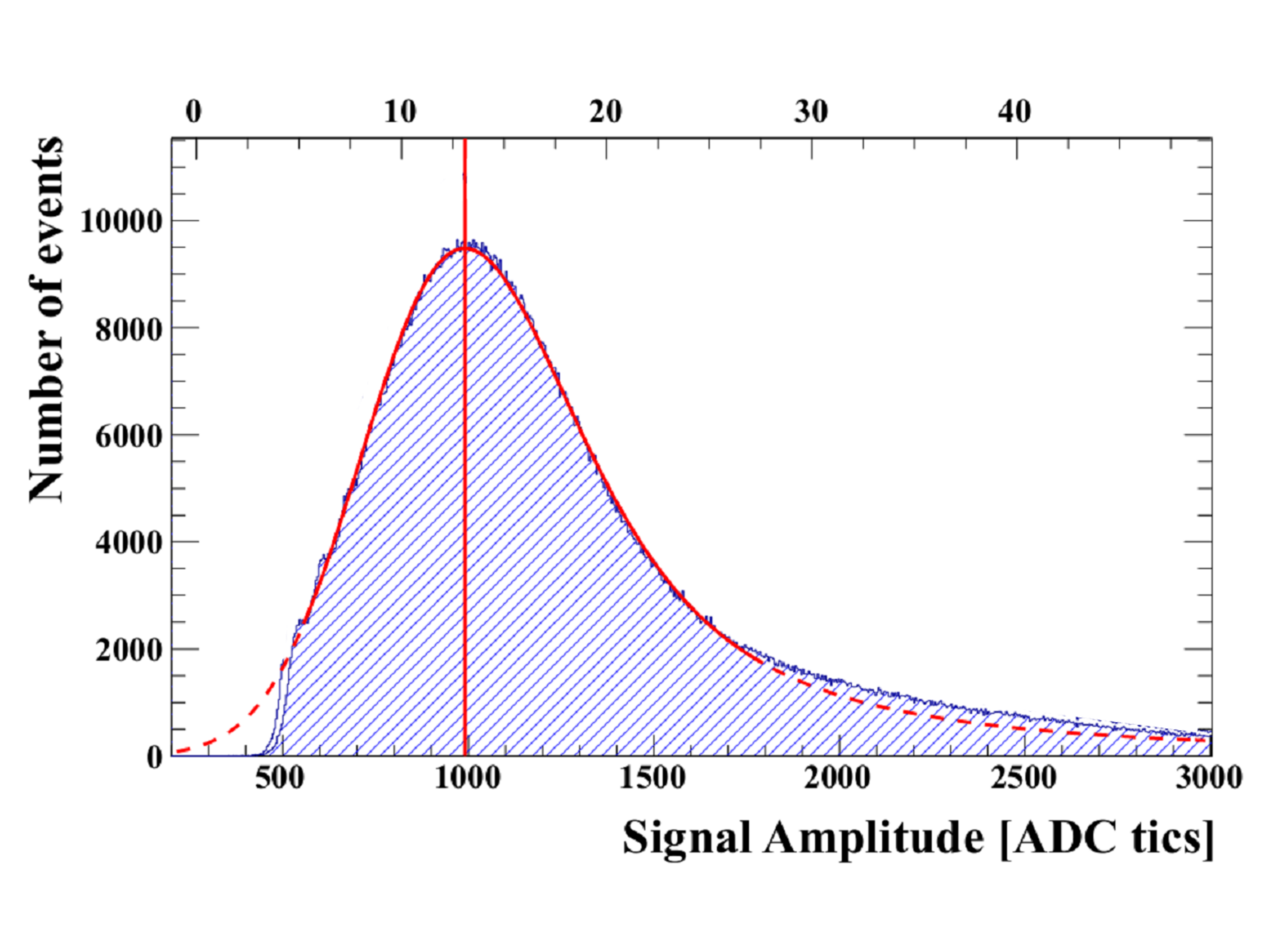}
\caption{Typical MIP spectrum obtained from a 2 GeV electron beam  in auto trigger mode, fitted with a Landau function convolved with a 
Gaussian function to determine the most probable value of the MIP~\cite{hartbri}.}
\label{fig_mip_plot}
\end{figure}

For that the SPIROC2b can run, in both auto trigger and external trigger mode. With an additionally external validation only events are saved which are coincident with an external signal (e.g. from a scintillator).  One result of the DESY testbeam is that the external validate mode is appropriate to suppress the SiPM noise and validate physics events in the detector.\\  
The measurements were done in high gain mode with an equalized cell gain. Figure~\ref{fig_mip_plot} shows a high statistics MIP spectrum. The measured ADC value from the front-end electronics is converted into the pixel scale by using the measured gain from single pixel spectra taken with LED light. The most probable value is around 11 pixels. Also the influence of the auto trigger threshold can be observed at the beginning of the spectrum. The spectrum is fitted with a Landau function convolved with a Gaussian function to determine the exact position of the most probable value. 

\section{Summary and outlook}

A new engineering prototype for an analog hadron calorimeter is currently being developed by the CALICE collaboration. The main goal is to demonstrate that a realistic LC detector with fully integrated front-end electronics can be built. The main challenge for the near future is to construct a full LC detector layer to test the signal integrity and the concept of power pulsing. The calorimeter base units were tested in the laboratory as well as in the DESY electron testbeam to characterize their features and to test the functionality and performance of all subcomponents as well as overall system aspects. Some of the recent measurement results are shown in this report, including some aspects of the LED calibration system, channel gain equalization and latest testbeam measurements. Furthermore, a layer setup has been used in a hadron testbeam environment for a system test and for measuring the time evolution of hadronic showers. The data analysis is ongoing. The next important steps are the construction of a small calorimeter stack ($\sim\,$10 layers) to investigate further the read out electronics, different scintillators and photodetector types, realistic mechanical structures and a scalable DAQ system.

\section*{Acknowledgment}
The author gratefully thanks Karsten Gadow, Erika Garutti, Peter G\"ottlicher, Oskar Hartbrich, Katja Kr\"uger, Sebastian Laurien, Marco Ramilli, Mathias Reinecke, Julian Sauer, Felix Sefkow, Mark Terwort and Sebastian Weber and the whole CALICE collaboration for very useful discussions and valuable contributions to the results presented here.


\begin{thebibliography}{1}

\bibitem{CALICE}
CALICE home page:\\ https://twiki.cern.ch/twiki/bin/view/CALICE/CaliceCollaboration

\bibitem{AdloffI}
C.~Adloff et al., \emph{Tests of a Particle Flow Algorithm with CALICE Test Beam Data}, JINST 6 (2011) P07005, arXiv: 1105.3417.

\bibitem{AdloffII}
C.~Adloff et al., \emph{Construction and commissioning of the CALICE analog hadron calorimeter prototype}, JINST 5 (2010) P05004, arXiv: 1003.2662.

\bibitem{Reinecke}
M.~Reinecke, \emph{Towards a full scale prototype of the CALICE tile hadron calorimeter}, Proc. 2011 IEEE Nuclear Science Symposium (NSS11); M.~Terwort, \emph{Concept and Status of the CALICE analog hadron calorimeter engineering prototype}, Proc. 2011 Technology and Instrumentation in Particle Physics (TIPP) arXiv:1109.0927; M.~Terwort and O.~Hartbrich for the CALICE Collaroration, \emph{Recent Advances of the Engineering Prototype of the CALICE Analog Hadron Calorimeter}, Proc. 2011 LCWS, arXiv:1201.5264.

\bibitem{Raux}
L.~Raux et al., \emph{SPIROC Measurement: Silicon Photomultiplier Integrated Readout Chips for ILC}, Proc. 2008 IEEE Nuclear Science Symposium (NSS08), NSSMIC.2009.5401891; M.Bouchel et al., \emph{SPIROC (SiPM Integrated Read-Out Chip): dedicated very front-end electronics for an ILC prototype hadronic calorimeter with SiPM read-out}, Proc. 2012 TWEPP, JINST 6(2011) C01098.

\bibitem{hartbri}
O.~Hartbrich, \emph{Commissioning and LED System Tests of the Engineering Prototype of the Analog Hadronic Calorimeter of the CALICE Collaboration}, DESY-thesis-2012-040, ISSN 1435-8085.




\end{thebibliography}
\end{document}